\documentstyle[12pt]{article}

\catcode`\@=11
\long\def\@makefntext#1{
\protect\noindent \hbox to 3.2pt {\hskip-.9pt  
$^{{\ninerm\@thefnmark}}$\hfil}#1\hfill}		

\def\@makefnmark{\hbox to 0pt{$^{\@thefnmark}$\hss}}  
	
\def\ps@myheadings{\let\@mkboth\@gobbletwo
\def\@oddhead{\hbox{}
\rightmark\hfil\ninerm\thepage}   
\def\@oddfoot{}\def\@evenhead{\ninerm\thepage\hfil
\leftmark\hbox{}}\def\@evenfoot{}
\def\sectionmark##1{}\def\subsectionmark##1{}}

\renewcommand{\thefootnote}{\fnsymbol{footnote}}

\newcounter{sectionc}\newcounter{subsectionc}\newcounter{subsubsectionc}
\renewcommand{\section}[1] {\vspace*{0.6cm}\addtocounter{sectionc}{1} 
\setcounter{subsectionc}{0}\setcounter{subsubsectionc}{0}\noindent 
	{\normalsize\bf\thesectionc. #1}\par\vspace*{0.4cm}}
\renewcommand{\subsection}[1] {\vspace*{0.6cm}\addtocounter{subsectionc}{1} 
	\setcounter{subsubsectionc}{0}\noindent 
	{\normalsize\it\thesectionc.\thesubsectionc. #1}\par\vspace*{0.4cm}}
\renewcommand{\subsubsection}[1]
{\vspace*{0.6cm}\addtocounter{subsubsectionc}{1}
	\noindent {\normalsize\rm\thesectionc.\thesubsectionc.\thesubsubsectionc. 
	#1}\par\vspace*{0.4cm}}

\newcounter{appendixc}
\newcounter{subappendixc}[appendixc]
\newcounter{subsubappendixc}[subappendixc]

\renewcommand{\appendix}[1] {\vspace*{0.6cm}
        \refstepcounter{appendixc}
        \setcounter{figure}{0}
        \setcounter{table}{0}
        \setcounter{equation}{0}
        \renewcommand{\thefigure}{\Alph{appendixc}.\arabic{figure}}
        \renewcommand{\thetable}{\Alph{appendixc}.\arabic{table}}
        \renewcommand{\theappendixc}{\Alph{appendixc}}
        \renewcommand{\theequation}{\Alph{appendixc}.\arabic{equation}}
        \noindent{\bf Appendix \theappendixc #1}\par\vspace*{0.4cm}}

\def\abstracts#1{{
	\centering{\begin{minipage}{12.2truecm}\footnotesize\baselineskip=12pt\noindent
	\centerline{\footnotesize ABSTRACT}\vspace*{0.3cm}
	\parindent=0pt #1
	\end{minipage}}\par}} 

\renewenvironment{thebibliography}[1]
	{\begin{list}{\arabic{enumi}.}
	{\usecounter{enumi}\setlength{\parsep}{0pt}
\setlength{\leftmargin 1.25cm}{\rightmargin 0pt}
	 \setlength{\itemsep}{0pt} \settowidth
	{\labelwidth}{#1.}\sloppy}}{\end{list}}

\newcounter{itemlistc}
\newcounter{romanlistc}
\newcounter{alphlistc}
\newcounter{arabiclistc}

\newcommand{\fcaption}[1]{
        \refstepcounter{figure}
        \setbox\@tempboxa = \hbox{\footnotesize Fig.~\thefigure. #1}
        \ifdim \wd\@tempboxa > 6in
           {\begin{center}
        \parbox{6in}{\footnotesize\baselineskip=12pt Fig.~\thefigure. #1}
            \end{center}}
        \else
             {\begin{center}
             {\footnotesize Fig.~\thefigure. #1}
              \end{center}}
        \fi}

\newcommand{\tcaption}[1]{
        \refstepcounter{table}
        \setbox\@tempboxa = \hbox{\footnotesize Table~\thetable. #1}
        \ifdim \wd\@tempboxa > 6in
           {\begin{center}
        \parbox{6in}{\footnotesize\baselineskip=12pt Table~\thetable. #1}
            \end{center}}
        \else
             {\begin{center}
             {\footnotesize Table~\thetable. #1}
              \end{center}}
        \fi}

\def\@citex[#1]#2{\if@filesw\immediate\write\@auxout
	{\string\citation{#2}}\fi
\def\@citea{}\@cite{\@for\@citeb:=#2\do
	{\@citea\def\@citea{,}\@ifundefined
	{b@\@citeb}{{\bf ?}\@warning
	{Citation `\@citeb' on page \thepage \space undefined}}
	{\csname b@\@citeb\endcsname}}}{#1}}

\newif\if@cghi
\def\cite{\@cghitrue\@ifnextchar [{\@tempswatrue
	\@citex}{\@tempswafalse\@citex[]}}
\def\citelow{\@cghifalse\@ifnextchar [{\@tempswatrue
	\@citex}{\@tempswafalse\@citex[]}}
\def\@cite#1#2{{$\null^{#1}$\if@tempswa\typeout
	{IJCGA warning: optional citation argument 
	ignored: `#2'} \fi}}

 1
 1
 1

\font\ninerm=cmr9

\begin{document}

\baselineskip=22pt
\centerline{\normalsize\bf SUMMARY OF THE INTERNATIONAL SYMPOSIUM ON HEAVY}
\baselineskip=16pt
\centerline{\normalsize\bf FLAVOR AND ELECTROWEAK THEORY}

\centerline{\footnotesize R. D. Peccei}
\baselineskip=13pt
\centerline{\footnotesize\it Department of Physics \& Astronomy, University of California}
\baselineskip=12pt
\centerline{\footnotesize\it Los Angeles, CA 90095-1547, USA}
\centerline{\footnotesize E-mail: peccei@physics.ucla.edu}
\vspace*{0.3cm}


\vspace*{0.9cm}
\abstracts{This summary discusses some of the topics which were of overarching 
interest at the Symposium.  These included, corrections to perturbative QCD
predictions; heavy quark physics; electroweak symmetry breaking; and physics
of the top quark.}
 
\normalsize\baselineskip=15pt
\setcounter{footnote}{0}
\renewcommand{\thefootnote}{\alph{footnote}}
\section{Introductory Remarks.}
Even with the best of will, it is impossible to summarize in 40 minutes
the 30 talks given
at the Symposium.  Thus, instead I will try to concentrate on a few topics
of overarching interest.  These included, corrrections to perturbative QCD
predictions; heavy quark physics; electroweak symmetry breaking; and physics
of the top quark.  There were many other interesting topics discussed at
the Symposium [perturbation theory resummation; renormalons; CP and 
automorphisms; mass shifts in strong magnetic fields; symmetry pattern of
mass matrices; etc.] which I, unfortunately, cannot properly cover in this
summary.  I apologize for this and refer the interested reader to the
appropriate contributions in these Proceedings.

\section{Corrections to Perturbative QCD.}
One of the recurring themes in the Symposium was that perturbative QCD
has its limitations.  Perturbative QCD gives accurate predictions as
long as the expansion parameter for the process in question is really
$\alpha_s$.  However, when this is not really the case, to obtain reliable
predictions, one must include corrections which depend in detail on the
physics of the problem.  Three examples were discussed at the Symposium,
each of which illustrated a particular way in which the relevant physics
dictated how to augment the perturbative QCD calculations.  Brodsky\cite{Brodsky} considered threshold effects in heavy quark production
in $e^+e^-$ collisions; Berger\cite{Berger} discussed resumming initial state
bremsstrahlung in top production at hadronic machines; and Wise\cite{Wise}
explained the role that color octet contributions have in hadronic production
of charmonia.  In each of these examples the underlying physics which causes
modifications to perturbative QCD is quite clear.  Indeed, for the processes
discussed by Brodsky and Berger analogous phenomena occur also in QED.
Nevertheless, each of these examples is a challenging area for QCD, if one
wants accurate predictions to compare with experiment.

Threshold production of pairs of charged fermions is sensistive to Coulomb
exchange.  For $e^+e^-\to \tau^+\tau^-$ near $\tau$-threshold one must
include the multiple Coulomb rescattering of the produced pairs.  Similarly,
for heavy quark-antiquark production for $\beta=\sqrt{1-4m^2_Q/s}~\to 0$ one 
must take into account of the gluonic Coulomb rescattering.  For both QED
and QCD one incorporates these effects through the introduction of a
Coulomb factor, which sums up the multiple exchanges of photons or gluons:
\[
S(x)=\frac{x}{1-e^{-x}}
\]
with
\[
x=\left\{ \begin{array}{ll}
\frac{\pi}{\beta}\alpha & \rm{QED}\\
\frac{\pi}{\beta} \frac{4}{3} \alpha_s & {\rm QCD}
\end{array}
\right.
\]
This Coulomb factor modifies the angular distribution at threshold, so that
the coefficient of $\cos^2\theta$ is not simply $\beta^2$ but $\beta^2S(x)$:
\[
\frac{d\sigma}{d\Omega} \sim
[2-\beta^2+\beta^2S(x) \cos^2 \theta]~.
\]
What Brodsky points out is that when one does the summing of the Coulomb
exchanges at threshold properly, one obtains in this Coulomb factor the running
coupling responsible for the binding of quarkonia $\alpha_V$, since the
same physics is involved.  Thus for QCD really one has
\[
x=\frac{4\pi}{3} \frac{\alpha_V(\beta^2s)}{\beta}~.
\]
This being the case, it may be possible to extract the coupling responsible
for the charmonium bound state spectrum by studying the threshold angular
distribution for $e^+e^-\to c\bar c$.  A real question, however, is if this
angular distribution is reflected faithfully in the angular distribution
of the corresponding charmed hadrons, or whether hadronization effects mask
entirely the Coulomb rescattering physics.

Berger\cite{Berger} discussed another example where to properly calculate the
physics of the problem one again has to sum up the effects of soft gluons--in
his case, radiated gluons form the initial state.  At the Tevatron the
production of top quarks comes dominantly from the process
$q\bar q\to t\bar t$.  In contrast, at the LHC this will occur mostly through
gluon fusion.  In the usual fashion, the hadronic cross section for top
production is then given by the convolution of the parton cross section and
the quark and antiquark distribution functions
\[
\sigma_{t\bar t}(s) = \int dx_1dx_2~
q(x_1)\bar q(x_2)~\hat\sigma_{t\bar t}(x_1x_2\hat s)
\]
The partonic cross section $\hat\sigma_{t\bar t}$ is known to $O(\alpha^2_s)$.
However, near threshold there are large corrections arising from the
bremsstrahlung of a soft gluon ($p_g\to 0$) from the initial state quarks
or antiquarks.  The single bremsstrahlung of a gluon introduces a factor
\[
sb=\int^1_0 dz[1+2\alpha_s \ln(1-z)]
\]
which, although finite, is large due to the soft gluon contribution at
$z\to 1$ ($p_g\to 0$ corresponds to $z\to 1$).  Thus, one should really
consider also multiple soft gluon emission.  As Berger discusses, one can
actually resum the bremsstrahlung logarithms $(\alpha_s\ln(1-z))^n$ from
multiple gluon emission and eventually one obtains a full enhancement factor
of the form\cite{sum}
\[
E\sim \alpha_s((1-z)^{2/3} m^2)\ln^2(1-z)~.
\]
However, from the above formula 
one sees that as $z\to 1$ one gets into scale values of $\alpha_s$
which are no longer in the perturbative regime.

There are different approaches of how to handle this.  For instance, in
this Symposium Berger\cite{Berger} discussed how one can use a principal
value regularization prescription 
to estimate the infrared sensitive part of the enhancement factor.  However, the important message is that, because
of these threshold effects, there is a bit of the top cross section at the
Tevatron that is {\bf uncalculable} in perturbative QCD.  In fact, as Berger
reported, what he and Contapanagos\cite{BC} do is to effectively set the
resummed contribution to zero for $\eta<0.005$ in the partonic cross section
because they cannot trust the answer below this value.  
They obtain in this way for the top cross section
at $\sqrt{s}=1.8~\rm{TeV}$, assuming $m_t=175~{\rm GeV}$, the value
\[
\sigma_{t\bar t}(1.8~{\rm TeV})=(5.5\pm 0.3)~ {\rm pb}~.
\]
Here the error is an estimate of the uncertainty coming from the structure
functions and the scale uncertainties.  Because the resummed contribution
contributes about 0.5 pb to the top cross section, the error coming from the
excluded region near $\eta=0$ probably is not significant.  Nevertheless, it
would be nice to have an estimate also of its possible magnitude.

Wise\cite{Wise} discussed some aspects of charmonium production in hadronic
collisions.  This is a topic of considerable interest since recent data at the
Tevatron showed that the production of $\psi,\psi^\prime$ and $\Upsilon$ is much
larger than was expected from a perturbative QCD quarkonia calculation\cite{charmpro}.  Schematically, quarkonium production is given by
convoluting the partonic cross-section for producing gluons of a certain fractional momentum with the gluon fragmentation function for quarkonia:
\[
d\sigma(p)=\int dz~\hat\sigma(z) P(p/z) D_{g\to Q\bar Q}(z)
\]
A naive estimate of the gluon fragmentation function can be obtained by
considering the same graphs which contribute to quarkonium decay.  This gives
for states whose decay involve two gluons
\[
D_{g\to Q\bar Q}(z) \sim \frac{\alpha_s^2}{\pi}
\frac{|\psi|^2}{m_Q^2} f(z) \sim\alpha_s^2 v^{3+2L}~.
\]
Here $v$ is the relative velocity of the bound quarks and $L$ is the angular
momentum associated with the produced quarkonia.

In his talk, Wise\cite{Wise} emphasized that because one is dealing with bound
state production one cannot just naively apply the same ideas that hold in
quarkonium decay.  Thus, for example, for the $L=1$ $\chi$-states besides
the naive result for $D_{g\to\chi}\sim \alpha_s^2v^5$, one can imagine\cite{octet} also production via an $L=0$ color octet intermediate state
which then decays via soft gluon emission to the $\chi$.  Such a color octet
contribution still involves a factor of $v^5$ but now is proportional to $\alpha_s$ not $\alpha_s^2$
\[
D^8_{g\to\chi} \sim \alpha_s v^5
\]
and hence, in principle, can give a much larger contribution.  Similar
considerations hold for $\psi$ production, where the naive quarkonium estimate
gives for the production of the $L=0~c\bar c$ state
\[
D_{g\to\psi}\sim \alpha_s^3 v^3~,
\]
while the contribution arising from an $L=0$ color octet intermediate state,
which then decays into a $\psi$ by emitting two soft gluons\cite{octet}, gives
\[
D^8_{g\to\psi}\sim \alpha_s v^7~.
\]
One gains a factor of $\alpha_s^2$ but at the price of a $v^4$ factor.  So here
it is not so clear whether the color octet contribution can give an enhancement.

Because detailed bound state calculations are not simple to do, it is difficult
to estimate reliably how much each of the above mechanisms really contributes
to the gluon fragmentation function into quarkonia.  Thus, it might be very
useful to have a diagnostic test which may help distinguish among these 
different mechanism.  Wise\cite{Wise} suggested one such diagnostic in his talk,
involving the alignment of the produced quarkonia.  If the color octet $L=0$
contribution dominates in $\psi$ production then, since the soft glouns are
irrelevant in the decay, one expects that the produced $\psi$ should be
transversally aligned.  Hence the produced leptons from the decay
$\psi\to \ell^+\ell^-$ should have an angular distribution proportional to
$1+\cos^2\theta$.  Unfortunately, the practical situation is not so simple since
about 30\% of the $\psi$'s come from radiative decays of produced $\chi$'s
($\chi\to\gamma\psi$) and so this dilutes the purity of the signal.  Furthermore, detecting the asymmetry in the production angle is hard experimentally for $\psi$'s produced at large transverse momentum, due to the
substantial kinematical boost of the produced leptons.

Still within QCD, but now in the non-perturbative sector, we heard also of some
nice work in the Symposium connected with novel quarkonia, like $B_c$ and 
baryons containing two different heavy quarks $QQ^\prime q$.  If one has systems
like $B_c$ or $QQ^\prime q$ with two heavy quarks of quite different masses, then mass effects can lead to substantial differences.  For instance, as
Chang\cite{CHChang} and Oakes\cite{Oakes} discussed, the hyperfine splitting
between $^3S_1$ and $^1S_0$ in the $B_c$ system is only about 70 MeV compared to
125 MeV in charmonium and 100 MeV in bottomonium.  For the double heavy baryons, one approach discussed by Chang\cite{CHChang} is to consider them as
bound states of a heavy diquark--light quark system:
\[
QQ^\prime q\sim \bar 3_{QQ^\prime} q~.
\]
This system is then not that dissimilar from a heavy-light meson, like $B_c$.
However, the diquark ($bc$) is much less tightly bound than the meson ($b\bar c$) \cite{CHChang}, with
\[
M_{bc} \simeq 6.6~{\rm GeV}~~{\rm versus}~~
M_{B_c\sim b\bar c} \simeq 6.3~{\rm GeV}~.
\]

\section{Heavy Flavor Decays.}
The physics of heavy quark systems is an important testing ground for our
theoretical understanding of QCD and of the electroweak interactions.  In
addition, heavy quark decays offer the opportunity for exploring further the
still poorly understood phenomena of CP violation.  The activity in this field,
which was mirrored in this meeting, roughly splits into two pieces:
\begin{description}
\item{i)} Improvements and refinements in dynamical calculations of weak
decay matrix elements by a variety of techniques: parton/quark models; chiral
perturbation theory; $1/N_c$ methods; lattice calculations; and QCD sum rules.
\item{ii)} Exploration of areas where one can probe better the standard model,
or look for signs of new physics.  These included CP violation in charged-$B$
decays; new ways to determine the angles in the unitarity triangle; studies of
non-CKM CP-violating phases; and the physics of $\tau$ lepton decays.
\end{description}

The talks of T. Huang\cite{THuang} and W. Bardeen\cite{Bardeen} in this 
Symposium provided two examples of attempts at better estimating dynamical
parameters in weak decays which are of considerable phenomenological interest.
Huang\cite{THuang} discussed SU(3) breaking effects for the predictions of
various quantities obtained by using heavy quark effective theory (HQET), using
QCD sum rules as a tool.  His results are as follows:
\begin{description}
\item{i)} The ratios of weak decay constants receive about a 10\% SU(3) breaking
corrections
\[
\frac{f_{Bs}}{f_{Bd}} = 1.18 \pm 0.05; ~~~
\frac{f_{Ds}}{f_D} = 1.13 \pm 0.03~.
\]
These results are quite compatible with lattice calculations.  Furthermore,
as Oakes\cite{Oakes} pointed out, the double ratio of the above quantities
is quite insensitive to SU(3) breaking.  These results are important for
phenomenology since, for example, the $B_s-\bar B_s$ mass difference $\Delta m_s$ can be derived from the $B_d-\bar B_d$ mass difference and CKM 
parameters once $f_{B_s}/f_{B_d}$ is known.
\item{ii)} The Isgur-Wise function and the operators coefficients of the HQET Lagrangian are quite insensitive to SU(3) breaking, with corrections of order a
few percent.  However, Huang\cite{THuang} finds that the slope parameters in
the Isgur-Wise function obey $\rho_s^2>\rho^2_{u,d}$, which is the opposite
behavior of that obtained in chiral perturbation theory.
\end{description}

Bardeen\cite{Bardeen} discussed another parameter of phenomenological 
importance for $B$ physics, the, so-called, bag constant $B_{Bd}$ which gives
a measure of the $\Delta B=2$ matrix element:
\[
\langle B_d|\bar d\gamma_\mu(1-\gamma_5) b\bar d\gamma^\mu(1-\gamma_5)
b|\bar B_d\rangle = \frac{8}{3} f_{Bd} M^2_{Bd} B_{Bd}~.
\]
Because this matrix element enters in the expression for the $B_d-\bar B_d$
mass difference, changes in the value of $B_{Bd}$ affect the constraints one
obtains for the CKM parameters obtained from the experimental value of this mass
difference.  Both lattice methods and QCD sum rules give values for
$B_{Bd}$ very close to unity.  Bardeen calculates this quantity
using $1/N_c$ methods.

The leading contribution for $B_{Bd}$ in a large $N_c$ expansion corresponds
to introducing the vacuum state in the above matrix element 
and leads to
$B_{Bd}=3/4$.  Non-leading contributions come from the connected matrix
elements involving the 2-current correlation
\[
{\rm corr}~= \int d^4q
\langle B_d|J_\mu(q) J^\mu(-q)|\bar B_d\rangle
\]
To proceed, Bardeen\cite{Bardeen} uses different techniques to evaluate the
above integral in different regions of momentum $q$, matching these calculations
at their interface.  Writing $q^\mu=m_bv^\mu+k^\mu$, Bardeen\cite{Bardeen}
uses HQET to calculate for $\Lambda_{\rm QCD} < k$, but uses an effective meson
theory for $k< \Lambda_{\rm QCD}$.

Both the HQET and the effective meson theory give integrals for the 
correction factor which are both infrared and ultraviolet sensitive and
matching these contributions gives two conditions.  One of them is a matching
scale which turns out to be $\lambda\simeq 600\sqrt{\alpha_s}~ {\rm MeV}$.
The other is a condition on the coupling strength in the effective theory and
Bardeen obtains $g^2=1/3$.  Remarkably, because of this second matching
condition, the result for $B_{Bd}$ that Bardeen\cite{Bardeen} obtains is
unaffected by the nonleading corrections in $1/N_c$:
\[
B_{Bd}=\frac{3}{4}[1-0.1(1-3g^2)]\longrightarrow \frac{3}{4}~.
\]
As Bardeen points out, it is not clear how general this result is.  For
instance, in his effective meson calculation he has included $B_d^*$ states
but not, for instance, $B_d^{**}$ states.  The inclusion of these further
states could change the coupling 
strength matching condition and thus the result for
$B_{Bd}$.  Nevertheless, it is troubling that there appears to be a
discrepancy between the value obtained for $B_{Bd}$ in lattice and QCD sum
rules calculations and in this $1/N_c$ calculation.

In the Symposium Lam\cite{Lam} also discussed the large $N_c$ limit, but 
applied to baryons which in this limit are just large collections of quarks:
$B\sim N_cq$.  As $N_c\to \infty$ these states are necessarily heavy, if the
quarks carry any mass.  Lam described in particular how to reconcile, in a
special kinematical limit, the fact that baryonic decays to $n$ mesons are highly suppressed in the large $N_c$ limit, with
\[
A(B\to B^\prime nM) \sim O\left( N_c^{\frac{2-n}{2}}\right)~,
\]
while individual Feynamn graphs are all of $O(N_c^{n/2})$ and, apparently,
grow with $N_c$.  The reconciliation is effected by having an infinite tower
of resonances in the theory in the large $N_c$ limit, with all the $MBB^*$
couplings being appropriately related.

Also somewhat theoretical was the nice discussion of C.-S. Huang\cite{CSHuang}
of how to recover the results of HQET in a Bethe-Salpeter formalism.  One
expects this to emerge in an analogous way that one recovers in the 
non-relativistic limit the Schr\"odinger equation from the Bethe-Salpeter
equation.  Nevertheless, it was nice to see how this obtains in detail, 
recovering both the spin symmetry as $M_Q\to \infty$ (provided one has vector
or scalar kernels) and the HQET form of the $1/M_q$ corrections.

Huang\cite{CSHuang} applied this covariant formalism to a model calculation of
exclusive semileptonic decays, where he extracted the Isgur-Wise function, and
to other heavy quark non-leptonic decays, like $D^*\to D\pi$.  Similar
calculation to these were discussed at the Symposium by C.-S. Kim\cite{Kim}, who
used a parton model for his calculations, and by L.-H. Chan\cite{Chan} who used
an effective low-energy Lagrangian similar to that discussed by Bardeen\cite{Bardeen}.

Kamal\cite{Kamal} also presented a model investigation, in his case concerning 
the color suppressed decays of the $B$ mesons into $\psi K$ and $\psi K^*$.
Kamal remarked that the usual calculation, where one drops the color pieces in
the effective Lagrangian after Fierzing the currents
 and where one uses
factorization, cannot reproduce the experimental values for either the ratio of
these modes or the polarization in the $\psi K^*$ mode:
\[
R=\frac{BR(B\to \psi K^*)}{BR(B\to\psi K)} =
1.71\pm 0.34; ~~~~
P_L(B\to\psi K^*) = 0.78\pm 0.07~.
\]
These two assumptions (using $N_c=3$) give a small $a_2$ amplitude, with
\[
a_2=c_2+\frac{1}{N_c} c_1\simeq 0.1
\]
What Kamal\cite{Kamal} pointed out was that everything works out--both here and
in color suppressed $D$-decays--if there is about a 10\% non-factorizable
contribution and an analogous $O(10\%)$ contribution from the color pieces in
the effective Lagrangian.  These contributions, effectively, conspire to change
the $a_2$ amplitude to a new effective amplitude, with
\[
a_2^{\rm eff} \simeq c_2~.
\]
So Kamal's results are similar to just imagining dropping the $1/N_c$ contributions--a suggestion made earlier in the literature\cite{N}.

Much more model-independent was the discussion of Paschos\cite{Paschos} at the
Symposium of inclusive semileptonic $B$-decays.  Because one is summing over
all hadronic final states, the inclusive rate can be written in terms of a
current commutator taken between $B$ states:
\[
W_{\mu\nu} = \int d^4x e^{-iqx}
\langle B|[J_\mu(x), J_\nu(0)]|B\rangle~,
\]
where $q^\mu$ is the momentum transfer to the final lepton pair.  This
quantity can be calculated in a controlled way for most of the allowed phase
space by using a combination of a light-cone expansion and HQET.  Thus, one
expects that the inclusive semileptonic rate should be reliably calculable
in terms of the parton model, augmented by the matrix elements of $O(1/m_b)$
operators [$D^2$ and $\sigma\cdot G$] arising from the light-cone expansion.
Unfortunately, these expectations are not realized in practice since the
experimental semileptonic branching ratio
\[
B_{sL} = \frac{\Gamma(B\to X\ell\nu_e)}{\Gamma(B\to ~{\rm all})} =
10.6 \pm 0.3
\]
is quite a bit smaller than the theoretical prediction of 12-13\%.

Paschos\cite{Paschos} discussed some possibilities for reconciling theory with
experiment.  This can happen readily if one, somehow, underestimated
the strength of the non-leptonic $B$-decays.  The favored idea here is that
the mode $b\to c\bar cs$ is underestimated.  However, to bring theory and
experiment in concordance one would need to boost up this mode so much that it
would lead to too much charm production ($N_c\sim 1.3$), in conflict with
observation.  It is possible that the discrepancy is the efffect of new
physics, where a favored effective operator is that given by
\[
L_{\rm eff} \sim \frac{1}{M^2_{\rm new}}
(\bar bs)_R (\bar qq)~.
\]
However, it may also just be that we, again, have failed to correctly
calculate the relevant non-leptonic matrix element.  History perhaps gives
credence to this last, more humble, hypothesis.  For kaons, the
$\Delta I=1/2$ enhancement is a factor of 20 which, even today, is only
partially understood.  We also have not really totally explained the factor
of 2 difference between the charm lifetimes, $\tau(D^+)/\tau(D_0) \sim 2$.  So
perhaps we should not be too concerned by a 20\% discrepancy in the
semileptonic $B$-decays!

D.-S. Du\cite{Du} in his talk at the Symposium suggested that one should
consider anew the possibility of having rather large CP-violating asymmetries
in charged $B$-decays.  This is an old suggestion\cite{Chau} which, however,
seems to be difficult to realize in practice.  To obtain a CP-violating
asymmetry in $B^\pm$-decays requires the interference of
two amplitudes with both
{\bf different} weak CP-violating phases and strong rescattering phases.
Although this occurs in practice, in general one of the amplitudes or one of
the phase differences is small and the net asymmetry is then also small.
Du\cite{Du} suggests that this may not happen for decays like $B^\pm\to \pi^\pm \pi^o$ where one is interfering a spectator decay amplitude with a
(space-like) Penguin amplitude.  Du gets a large effect by assuming that the
size of the space-like Penguin amplitude is related to the Brodsky-Lepage\cite{BL}
form factor:
\[
\langle \pi\pi|J|0\rangle \sim \frac{i\alpha_s}{M_B^2}~.
\]
This gives him an amplitude which is comparable in size to the spectator decay
amplitude and in which the rescattering phase is maximum.  Because the two
amplitudes in question involve $V_{ub}$ and $V_{td}$, respectively, the weak
phases are also comparable.  So, in principle, one could get large effects.
Unfortunately, it is difficult to judge how reliable the Penguin estimate of
Du\cite{Du} is.  At any rate, he has raised an interesting issue.

Tau decays were also discussed at the Symposium, both as a beautiful laboratory for applying current algebra and dispersion relation techniques\cite{Truong} and as a place to look for new physics\cite{Nelson}.  Truong\cite{Truong}
emphasized that the current algebra soft pion relation in the limit of
$p^\mu\to 0$:
\[
\langle B\pi|V_\mu|A\rangle = \frac{1}{f\pi}
\langle B|A_\mu|A\rangle~,
\]
when used with the Pad\'e techniques to build-in unitarity, can be very
powerful.  Indeed, by these means it is possible to make successful predictions
for multipion $\tau$-decays ($\tau\to n\pi\nu_\tau$), including resonance 
channels, like $\tau\to \pi\rho\nu_\tau$.  
Nelson\cite{Nelson}
instead concentrated on what limits on new physics could be obtained from
$\tau$-decays at the proposed  Beijing tau-charm factory.  He showed that, by
looking at the $\tau\to \rho\nu_\tau$ and $\tau\to A_1\nu_\tau$ decays and
analyzing the $\rho$ and $A_1$ polarization through their further decays, one
can obtain limits on the scale associated with new V-A interactions of the
$\tau$ which are of $O(\Lambda\sim 1~{\rm TeV})$.  Nelson\cite{Nelson} also
showed that one could test for possible CP-violating asymmetries in the
charged $\tau$-decays to quite a reasonable level.  For instance, writing the
amplitude for $\tau^\pm\to \rho^\pm\nu_\tau$ as $r^\pm = |r| e^{i\phi}$, at
a tau-charm factory one could hope to determine $\delta r/r$ to about 0.1\%
and $\delta\phi$ to about $1^\circ$.

\section{Electroweak Symmetry Breaking.}
The third subject of great interest at the Symposium was electroweak physics.
Here there are a few facts which were agreed by all the speakers, either
implicitly or explicitly:
\begin{description}
\item{i)} The standard model gives an amamzingly accurate description of
a large body of precise electroweak data\cite{Kang,Ellis}.  An example
being provided by the very accurate value of $sin^2\theta_{\rm eff}=
0.2315\pm 0.0004$.
\item{ii)} The physics underlying the breakdown of $SU(2)\times U(1)\to U(1)_{\rm em}$ occurs at scale of $O(1~{\rm TeV})$.
\item{iii)} The large mass of the top quark, with $m_t\sim O(v)$ and where
$v=(\sqrt{2}~G_F)^{-1/2}\simeq 250~{\rm GeV}$ is the scale associated with the
Higgs vev, is significant.  Although what exactly this is telling us is not yet
totally clear\cite{Parke,Young}.
\end{description}
The focus of the discussion at the Symposium was on the {\bf disputable aspects}
of the above points.  For example, are there hints of small discrepancies with
the standard model in the data?  Or, what really is the physics which is at the
root of the symmetry breakdown?  Or, what is the real significance of having
top so heavy?

Probably the central issue of particle physics today is what is the mechanism
which causes the $SU(2)\times U(1)$ breakdown.  Two camps exist.  Partisans
of the first camp believe that the breakdown is due to the vev of some
elementary scalar(s) field(s).\cite{Ellis}  This is the original mechanism
suggested for the spontaneous breakdown of the standard model.  However, to make
this mechanism natural the belief now is that one needs to have also some
supersymmetry which survives to low energy.  Partisans of the second camp
believe instead that the spontaneous breakdown of $SU(2)\times U(1)$ is due
to the formation of condensates of some underlying fermions\cite{Simmons}.
That is, the breakdown of $SU(2)\times U(1)$ is dynamical.  It is possible that
what condenses to break the symmetry is just $\langle t\bar t\rangle$, but
generally it is assumed that the condensing fermions are fermions of a new
theory--technicolor.

If the first option above is the truth and one has some low energy supersymmetry,
then eventually one should see plenty of signals.  All known excitations will
have superpartners and their spectrum will inform us of how precisely the
supersymmetry is broken down in nature.  Furthermore, since to implement the
supersymmetry one needs at least 2 Higgs doublets, one should also observe
the scalar excitations connected with an extended Higgs sector.\cite{Ellis}
In general, a relatively light Higgs boson ($M_h \leq M_Z$) is symptomatic of
supersymmetry.  One knows from direct searches at LEP that the standard model
Higgs boson has a mass $M_H > 65~{\rm GeV}$.  As Ellis\cite{Ellis} discussed
at the Symposium, from indirect fits to precision electroweak data one infers
that $M_H=76^{+100}_{-50}~{\rm GeV}$.  Optimistically, he concluded that 
such a "light Higgs" perhaps is already a hint of supersymmetry.  Whether
this is so only time (and more data!) will tell.

The breakdown of the electroweak symmetry by a Higgs vev which is stabilized by
supersymmetry is, in many respects, a much "safer" option than dynamical symmetry breaking.  Principally this is because it does not tie the scale of
$SU(2)\times U(1)$ breaking to the physics scale responsible for generating the
Yukawa couplings of the Higgs to the fermions, which are responsible for fermion
masses.  This cannot be avoided when the symmetry breaking is dynamical and, in
these latter theories, one is forced to have the fermion mass  generation scale near to the $O({\rm TeV})$ scale of $SU(2)\times U(1)$ breaking.

Simmons\cite{Simmons} discussed at the Symposium how the large mass of the top
makes life even more difficult.  Typically, when the electroweak breakdown is
caused dynamically, one generates fermion masses though effective 4-fermion
interactions between the ordinary quark and leptons and a new set of fermions
(technifermions) whose condensation causes the breakdown.  This ETC mechanism\cite{ETC} provides an effective Lagrangian of the form
\[
L_{\rm eff}\sim\frac{1}{M^2}(\bar TT)(\bar\psi\psi)
\]
where $M$ is the scale of the ETC interactions which connect the ordinary
fermions $\psi$ with the technifermions $T$.  The breakdown of
$SU(2)\times U(1)$ occurs as a result of the formation of a 
$\langle \bar TT\rangle$ condensate.  Because of the above effective
interactions, these condensates also give mass to the ordinary fermions.  Since top has such a large mass and
\[
m_t\sim\frac{\langle \bar TT\rangle}{M^2}~,
\]
the fermion mass generating scale $M$ cannot be very large.\cite{Simmons}
Because the electroweak breaking scale associated with the 
$\langle \bar TT\rangle$ condensate is of $O({\rm TeV})$ [i.e.
$\langle\bar TT\rangle\sim ({\rm TeV})^3$] the scale $M\sim O(10~{\rm TeV})$,
at most.  The presence of such "low scales" for new physics associated with
fermion mass generation, in general, produces unwanted flavor changing neutral
currents and one must devise rather clever schemes\cite{walking} to avoid these
troubles.  Furthermore, the technicolor condensates themselves produce small
changes in the expectations of precision electroweak tests and these changes 
are not favored experimentally.  For instance, as Kang\cite{Kang} discussed,
the so-called $S$ parameter is, in general, positive as a result of having
$\langle\bar TT\rangle$ condensates, while data prefers $S<0$.

Simmons\cite{Simmons} pointed out an especially serious problem for classes of
ETC models precisely in the area where there appears to be some discrepancy
between the data and the standard model.\cite{Kang,Ellis}
This is in the ratios of the widths of the $Z$ into $b\bar b$ and $c\bar c$
states to the total width.  Experimentally, one has
\[
R_b=\frac{\Gamma(Z\to b\bar b)}{\Gamma(Z\to ~{\rm hadrons})} =
0.222 \pm 0.002; ~~~~
R_c=\frac{\Gamma(Z\to c\bar c)}{\Gamma(Z\to ~{\rm hadrons})} =
0.154 \pm 0.008,
\]
while the standard model expectations are centered around 0.216 and 0.173, 
respectively.  Simmons noted that for models where the ETC interactions commute
with $SU(2)$, then the same interactions which give a mass to the top also
give a specific shift to $R_b$, but no shift in $R_c$.  Unfortunately, these
models give a shift (of about 4\%) in the {\bf wrong} direction and therefore
are strongly disfavored by the data.  One can, however, invent models where
the ETC interactions and $SU(2)$ do not commute and change the sign of the
$R_b$ shift [essentially, one needs to change a $\vec\tau\cdot\vec\tau$
interaction to a $1\cdot 1$ interaction].  However, the resulting models are a bit recondite in that different families are treated differently and one may run into some problems with universality.

\section{Physics of Top.}
The discovery of the top quark at Fermilab\cite{top} was one of the year's
highlights.  The results of CDF and DO are as follows:
\[
\begin{array}{lll}
m_t = 178 \pm 11 \pm 9~{\rm GeV}; &
\sigma_{t\bar t}(m_t) = 6.8^{+3.6}_{-2.4}~{\rm pb} & {\rm (CDF)} \\
m_t = 199^{+19+14}_{-21-21}~{\rm GeV}; &
\sigma_{t\bar t}(m_t)=6.4 \pm 2.2~{\rm pb} & {\rm (DO)}
\end{array}
\]
At the Symposium the sensitivity of these results to possible new physics
contributions were discussed by Parke\cite{Parke} and C.-S. Li\cite{Li}, who
specifically considered the effects of possible supersymmetric corrections to
the top production cross section.  B.-L. Young\cite{Young} instead speculated
on possible non-standard couplings for the top, which may be more evident
because of its large mass.

Although speculation of new physics associated with the top is fair game, there
is already really not too much room to manuever.  For instance, the combined
value for the top mass coming from the CDF and DO measurements,
$\langle m_t\rangle=(181 \pm 12)~{\rm GeV}$ is actually in quite good agreement
with that obtained through the precision electroweak tests (when the Higgs mass
is considered a free parameter) reported by Ellis\cite{Ellis}:
$m_t=(155 \pm 14)~{\rm GeV}$.  The average of both these values gives a top
mass of $\langle \!\langle m_t\rangle \!\rangle = 172 \pm 10~{\rm GeV}$.  For
this mass the latest calculation of the top cross section reported by 
Berger\cite{Berger} here, of $\sigma_{t\bar t}(m_t)=5.5\pm 0.3~{\rm pb}$, is
in reasonable agreement with the CDF and DO values.  So, it could well be that
also for top everything is standard!

The discussion of Parke\cite{Parke} at the Symposium emphasized what physics
could explain possible disagreements between theory and experiment.  Although
he presented a more speculative interpretation for the present data, this
exercise is very useful nevertheless.  As usual, a good way to test
sensitivity to new physics is to introduce contact terms describing new
interactions of the top with the ordinary quarks, respecting the symmetries of
the standard model.  Parke\cite{Parke} discussed 4-fermion interactions of the
type
\[
L_{\rm eff} = \frac{g_3^2}{\Lambda^2_1} (\bar q1q)(\bar t1t) +
\frac{g_3^2}{\Lambda^2_8}(\bar q\lambda q)(\bar t\lambda t)
\]
and indicated that present data bounds the scales $\Lambda_1$ and $\Lambda_8$
to be above a TeV.  He also discussed more specific models, like the 
coloron model\cite{coloron} where the color $SU(3)$ group of QCD arises as
a result of the spontaneous breakdown of an $SU(3)\times SU(3)$ group.  The
octet of gauge bosons which acquire mass--the colorons-- have a mass
$M\sim\Lambda_8$ but have different couplings to ordinary quarks ($\sim\tan\theta$) than to top ($\sim\cot\theta$).  Such a coloron model\cite{coloron} predicts distinctive transverse momentum distortions for
top production and structure in the invariant mass of the produced $t\bar t$-pairs.\cite{Parke}

B.-L. Young\cite{Young} discussed another aspect of possible anomalies connected with top.  If the symmetry breakdown of the electroweak theory is
dynamical, it is natural to expect anomalous interactions of the Nambu-Goldstone
bosons with the fermions in the theory
\[
L_{\rm eff} \sim \frac{\kappa}{\Lambda} \bar\psi\gamma_\mu\psi
\partial^\mu\xi + \ldots
\]
By the equivalence theorem, discussed here by Y.-P. Kuang\cite{Kuang}, these
couplings eventually give rise to anomalous vertices of the fermions with the
gauge bosons.  For top these anomalous vertices could be of significant
strength, since one expects $\kappa\sim O(m_t/\Lambda)$ and $\Lambda$ to be in
the TeV range.  Therefore, because of the large mass of top, one could be
sensitive to new phenomena connected with the way the electroweak symmetry
breaks down.  These anomalous vertices, as Young\cite{Young} discussed, could be
responsible for the small discrepancy in $R_b$ and could also give rise to other
phenomena, like flavor changing decays of the top, which may be observable some
day.\cite{HPZ}

\section{Concluding Remarks.}
My conclusions are very simple.  This has been an exciting and fun meeting to
be at, with plenty of physics bubbling up!  Such a meeting would not have been
possible without all the hard work done by the Organizers.  On behalf of all
of the participants, I would like to thank them for their splendid
hospitality.

\section{Acknowledgments.}
This work was supported in part by the Department of Energy under Grant 
No. FG03-91ER40662.  I would like also to thank Tuo Huang for having made my
brief stay in Beijing so enjoyable.
\vspace{0.9 cm}

{\bf{References.}}

\end{document}